\documentclass[%
 reprint,
superscriptaddress,
 amsmath,amssymb,
 aps,
 prl,
]{revtex4-2}

\usepackage{graphicx}
\usepackage{dcolumn}
\usepackage{bm}
\graphicspath{{figures/}}
\usepackage{hyperref}

\begin{document}

\title{Bulk morphology of porous materials at sub-micrometer scale \\ studied by multi-modal X-ray imaging with Hartmann masks}

\author{Margarita Zakharova}
 \email{margarita.zakharova@partner.kit.edu}
\author{Andrey Mikhaylov}
\affiliation{%
 Institute of Microstructure Technology (IMT), Karlsruhe Institute of Technology (KIT), 76344 Eggenstein-Leopoldshafen, Germany
}%
\author{Stefan Reich}%
\affiliation{%
 Institute of Photon Science and Synchrotron Radiation (IPS), Karlsruhe Institute of Technology (KIT), 76344 Eggenstein-Leopoldshafen, Germany
}%
\affiliation{
 Current affiliation: Fraunhofer Institute for High-Speed Dynamics, Ernst-Mach-Institute, EMI, 79104 Freiburg, Germany
}%

\author{Anton Plech}
\affiliation{%
 Institute of Photon Science and Synchrotron Radiation (IPS), Karlsruhe Institute of Technology (KIT), 76344 Eggenstein-Leopoldshafen, Germany
}%

\author{Danays Kunka}
\email{danays.kunka@kit.edu}
\affiliation{%
 Institute of Microstructure Technology (IMT), Karlsruhe Institute of Technology (KIT), 76344 Eggenstein-Leopoldshafen, Germany
}%

\date{\today}

\begin{abstract}
We present the quantitative investigation of the submicron structure in the bulk of porous graphite by using the scattering signal in the multi-modal X-ray imaging with Hartmann masks. By scanning the correlation length and measuring the mask visibility reduction, we obtain average pore size, relative pore fraction, fractal dimension, and Hurst exponent of the structure. Profiting from the dimensionality of the mask, we apply the method to study pore size anisotropy. The measurements were performed in a simple and flexible imaging setup with relaxed requirements on beam coherence.
\end{abstract}
\maketitle
Porous materials are challenging objects for characterization: they typically exhibit a wide range of pore sizes, solid material opacity, possible anisotropy of the pores, and structure inhomogeneity in the bulk. Many conventional microscopic techniques have a limited field of view, which often makes the characterization they offer confined and incomprehensive. Excellent penetrating capabilities of X-ray radiation enable it to study otherwise opaque materials in a non-destructive way. 

Multi-modal X-ray imaging can offer a large field of view and provide different types of information retrieved from the projection of the sample (or a set thereof). Many X-ray imaging methods are based on the analysis of the changes of wavefront modulation relative to a reference image \cite{pfeiffer2006phase,vittoria2015x,wen2010single}. The reference image records the initial wavefront modulated by a chosen periodic optical element, such as a phase grating \cite{pfeiffer2006phase,yashiro2010origin}, a speckle filter \cite{berujon2012x,zanette2014speckle}, a Fresnel zone plate \cite{kagias20162d}, a lens array \cite{dos2018shack, reich2018scalable, mikhaylov2020shack}, or X-ray absorption masks, including Hartmann masks \cite{zakharova2019inverted, letzel2019time}. The disturbances in the wavefront modulation introduced by a chosen object are analyzed and attributed to its properties. The general decrease in the intensity is related to transmission contrast, the shift of the modulation to the differential phase contrast, and the dampening of the projected modulation to the scattering contrast. The latter arises from an ultra-small angle scattering of X-rays on the fine inhomogeneities in the sample and the phase signal below the resolution limit of the imaging setup \cite{wen2008spatial,yashiro2010origin, koenig2016origin}. 

The scattering signal can be evaluated in various ways, depending on the imaging method: through visibility reduction \cite{pfeiffer2006phase,yashiro2010origin}, an increased width of an individual modulation peak \cite{dos2018shack,vittoria2015x}, or the change in the first-order harmonic in the Fourier domain \cite{wen2008spatial,wen2010single}. It offers an opportunity to probe structures at the sub-$\mu m$ scale and retrieve information on their microscopic textural properties while performing macroscopic imaging.

\begin{figure}
\includegraphics[scale=1]{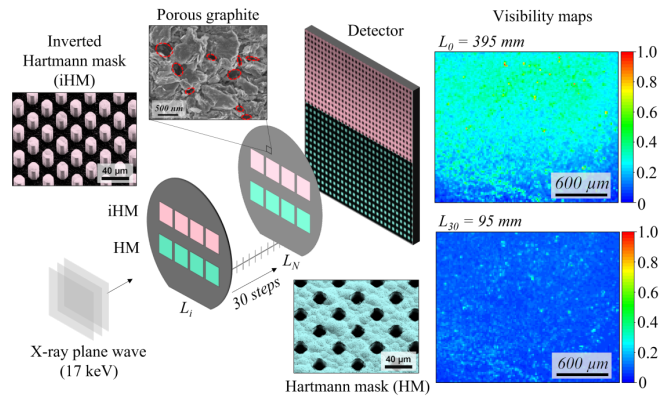}
\caption{\label{fig:fig1}Experimental setup for visibility measurements. Monochromatic X-rays are incident on the mask, which is moving towards the detector with steps of 10 mm. The measurements were done with inverted and conventional Hartmann masks of different periods produced on porous graphite (see Supplemental Material for Hartmann masks specifications). For each step, a projection is recorded from which the visibility map is plotted. The visibility maps for two distances are shown in the right part of the graph.}
\end{figure}

Hartmann masks are arrays of high-absorbing structures alternated by areas of high X-ray transmission. The two main designs are an array of holes (conventional Hartmann mask) and an array of absorbing pillars (inverted Hartmann mask) \cite{zakharova2019comparison}. Multi-modal X-ray imaging with Hartmann masks offers the advantages of setup robustness, relaxed requirements on the beam coherence, and versatility of the setup in the positioning of mask and sample (Fig.\ref{fig:fig1}). Hartmann masks can be fabricated by UV lithography combined with gold electroplating and scaled to the required field of view \cite{zakharova2018development}.

We used Hartmann masks of both designs to study the bulk morphology of porous graphite by analyzing the scattering contrast available through multi-modal X-ray imaging. Scattering contrast arises from the autocorrelation of electron density distribution, which peaks at a specific correlation length \cite{lynch2011interpretation}. The specific peak correlation length $\xi$ probed in multi-modal X-ray imaging depends on the setup parameters \cite{wen2008spatial, lynch2011interpretation, prade2016short}. For a fixed setup where the periodic optical element is placed before the object, it is defined as follows:
\begin{equation}
\xi=\frac{\lambda L}{P}
\label{eq:xi},
\end{equation}
where $\lambda$ is the wavelength of the X-ray radiation,  $P$ is the period of the wavefront modulation at the detector plane, and $L$ is the distance from the object to the detector. By varying any of the values in Eq.~(\ref{eq:xi}), one can perform a scan of the correlation length and determine the autocorrelation function for the object under study \cite{taphorn2020grating, prade2016short, kagias2021simultaneous}.

Imaging with Hartmann masks has no restrictions on the positioning of the object; therefore, it provides fine scanning of the correlation length in the sub-$\mu m$ range through the variation of $L$. By analyzing the visibility reduction, we can retrieve quantitative structural parameters of graphite using its real-space correlation function. This method can be applied to study textural properties of various complex microstructure systems, including \emph{in situ} and \emph{operando} measurements, and can be extended to laboratory setups.  

Visibility of the periodic modulation is a convolution of the modulation function and the scattering function. The modulation function is determined by the mask geometry and instrument resolution. When the visibility of the pattern decreases, it indicates stronger scattering by the object. The relationship between the decrease in visibility and the autocorrelation function of the electron density was defined elsewhere \cite{strobl2014general}. From this, we can determine the scattering intensity $S$ as \cite{yashiro2010origin, taphorn2020grating}: 
\begin{equation}
S=-\ln\left(\frac{V_{\xi}}{V_0}\right)=\sigma t(1-G\left(\xi\right)),
\label{eq:S}
\end{equation}
where $V_0$ is the visibility of the projected mask pattern at no scattering ($\xi$ = 0), $V_{\xi}$ is the visibility of the projected mask pattern at the correlation length $\xi$, $\sigma$ is the macroscopic scattering cross-section, $t$ the sample thickness, and $G(\xi)$ is the real-space autocorrelation function of electron density at the correlation length $\xi$ (Eq.~(\ref{eq:xi})).

The autocorrelation function depends on the structural properties of the object under study. For the dense but disordered structures such as graphite, the real-space autocorrelation function derived for random self-affine density distributions can be used  \cite{sheppard1996scattering,andersson2008analysis, andersson2008stress, hunter2006tissue}
\begin{equation}
G(\xi)=\frac{2}{\Gamma(H+1/2)}\left(\frac{\xi}{2a}\right)^{H+1/2} K_{H+1/2}\ \left(\frac{\xi}{a}\right)
\label{eq:G_b},
\end{equation}
where $a$ is characteristic size parameter, $K_{H+1/2}(x)$ the modified Bessel function of the second kind of real order $(H+1/2)$ and $\Gamma$ the Gamma function. $H$ is the so-called Hurst exponent ($0<H<1$) related to the dimensionality of the structure, namely to the interface roughness between the two phases of the material. 

The Hurst exponent $H$ from Eq.~\ref{eq:G_b} is determined by the space-filling capacity of the structure and defines its fractal dimension $D = D_E + 1 - H$, where $D_E$ is the Euclidean dimension of the scattering structure: 1 for filamentous, 2 for sheet-like, and 3 for bulk scatterers \cite{voss1985random, hunter2006tissue}. From this relation one can see that the value of the Hurst exponent reflects the fractal dimension and the specific surface area of porous material. Two domains are usually discussed: $H>1/2$ indicating that the density distribution is persistent (long-range correlations) with smoother and more interconnected pores, and $H<1/2$, which corresponds to antipersistent distributions with smaller and more confined pores and low permeability \cite{prostredny2019modelling}.  The characteristic size of the structure $d$ for random two-phase media  can be understood as the average pore size and is derived from the size parameter $a$ and the Hurst exponent $H$ as follows \cite{andersson2008stress}
\begin{equation}
d = \frac{2\sqrt\pi a\Gamma(H+1/2)}{\Gamma(H)}
\label{eq:d}.
\end{equation}

In our work, we also applied a simplified phenomenological fitting function \cite{sinha1988x, andersson2008analysis}  for the analysis of the pore size anisotropy
\begin{equation}
G(\xi)=\exp\left[-\left(\frac{\xi}{d}\right)^\alpha\right]
\label{eq:G_a},
\end{equation}
where $\alpha = D_E/2 + H$. Since $0 < H < 1$, the allowed range of values for $\alpha$ will depend on the Hurst exponent and the Euclidean dimension \cite{hunter2006tissue, sheppard1996scattering}. The allowed ranges of $\alpha$ will lay in the bounds $0.5 < \alpha_1 < 1.5$, $1 < \alpha_2 < 2$ and $1.5 < \alpha_3 < 2.5$ \cite{hunter2006tissue} for $D_E = (1, 2, 3)$, respectively. The value of $\alpha$ is related to the phase boundary and interface roughness \cite{sinha1988x, nesterets2008origins, andersson2008analysis}.

To ensure the high visibility of the mask pattern, minimize possible noise, and demonstrate the capabilities of the method, we fabricated Hartmann masks directly on the porous material. Nevertheless, such measurements can as well be performed with a setup where the Hartmann mask is manufactured on a separate low-absorbing wafer, for example, polyimide \cite{koch2015note}. 
The wafers purchased from Ohio Carbon Blank, Inc were synthetic graphite labeled "angstrofine" with a porous structure and an average grain size of 1 $\mu m$ as reported by the supplier (identifier EDM-AF5). The Hartmann masks were manufactured in a cleanroom environment using UV lithography, gold electroplating, and resist stripping \cite{zakharova2019comparison}. Inverted and conventional Hartmann masks of various periods (30, 40, 60, and 80 $\mu m$) with the gold height of about 30 $\mu m$ have been fabricated on the same substrate (see Supplemental Material). Additionally, a large area (5 cm x 5 cm) inverted Hartmann mask was produced on a different wafer from the same graphite plate. The masks will be further referred to as "Mask type-Period", e.g. iHM-30 for the inverted Hartmann mask of period 30 $\mu m$ and HM-30 for conventional Hartmann mask of period 30 $\mu m$.

Visibility measurements have been carried out at the IPS imaging cluster of the KIT synchrotron facility. For the measurements, a quasi-monochromatic beam with an energy of 17 keV and energy bandwidth of 2 \% was used. Detection of the X-rays was performed by an Andor Neo 5.5 camera imaging an X-ray scintillator (LuAG) by lens coupling (magnification of 2.73) to achieve an effective pixel size of 2.4 $\mu m$. The experimental setup is shown in Fig.~\ref{fig:fig1}. X-rays are incident on the HM or iHM, which are placed on a linear stage. The stage can move in the range of 300 mm. The masks on graphite were moving along the beam path, and projections were recorded for each step. In this way, we obtained the dependence of visibility on the position of the mask (distance to the detector). The mask was moved from 95 mm to 395 mm to the detector in 30 steps (step size 10 mm). An additional projection of the iHM-50 at a distance of 1120 mm from the detector was acquired. The changes in the visibility were attributed to the ultra small angle scattering in graphite, which was analyzed using the autocorrelation function for a random inhomogeneous two-phase media according to Eq.~\ref{eq:G_b}. The instrument resolution (source size) was much higher than the mask period, therefore the change of the modulation function with distance is neglected (see Supplemental Material for experimental setup details). The change in visibility is entirely attributed to the sample properties.
\begin{figure}
\includegraphics[scale=1]{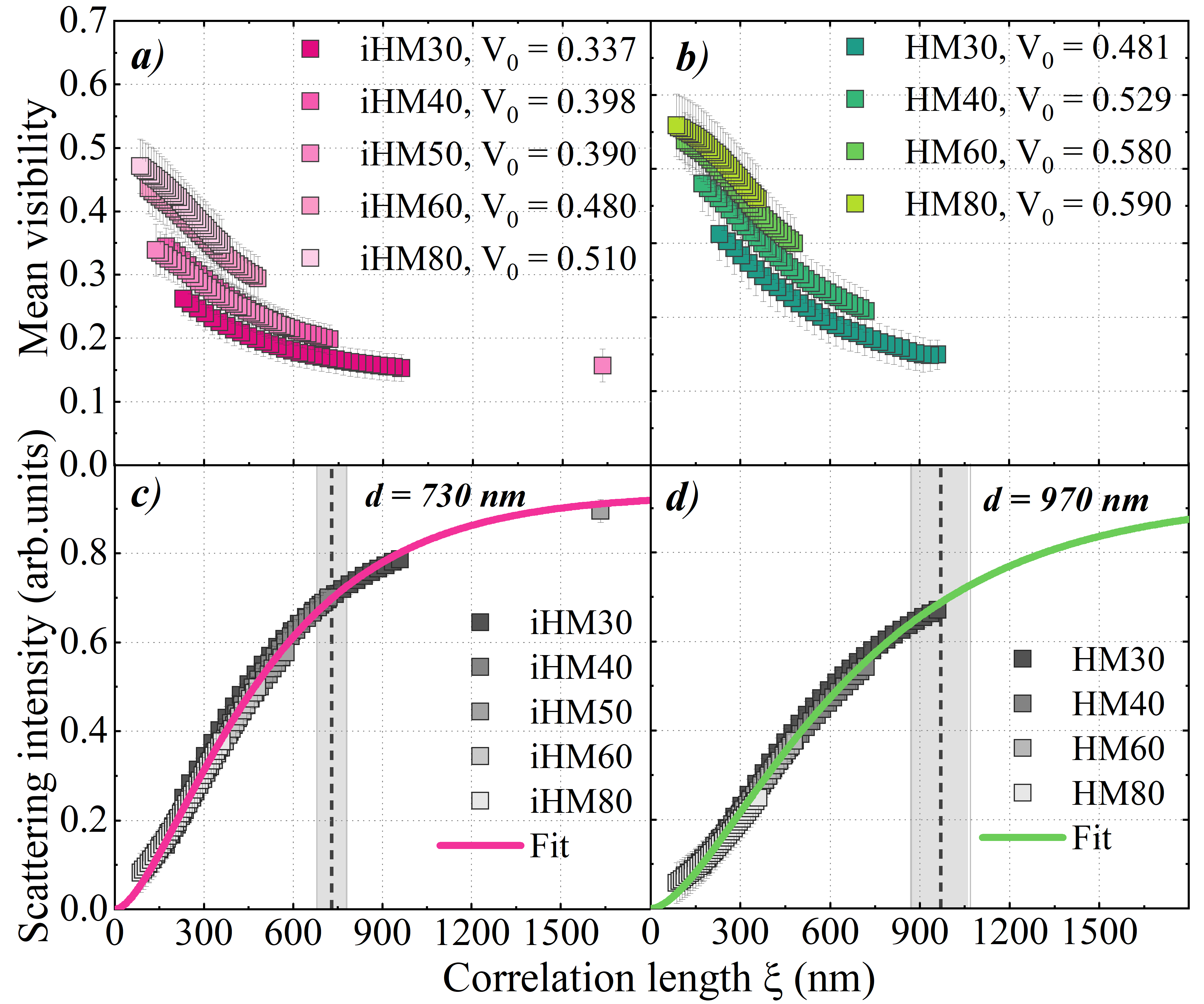}
\caption{\label{fig:fig2}Mean visibilities (a, b) and scattering intensities (c, d) (Eq.~\ref{eq:S}) for (a, c) inverted and (b, d) conventional Hartmann masks on graphite obtained for correlation lengths $\xi$ scanned in the experiment. The error bars are the standard deviations of the obtained signal.}
\end{figure}
The visibility maps were calculated by $V=(I_{max}-I_{min})/(I_{max}+I_{min})$, where $I_{max}$ is the maximum and $I_{min}$ the minimum intensity within a beamlet zone. Mean visibilities were calculated as mean values of the visibility maps in a selected pattern square. In Fig.~\ref{fig:fig2}(a,b) the mean values of the visibility maps are plotted for the different masks with error bars representing the standard deviation.  Assuming the parallel beam geometry, the visibility decreases as the correlation length value gets larger due to the dampening of the mask contrast at the larger mask-to-detector distances. 

To determine the scattering intensity (Eq.~\ref{eq:S}) we need to know the mean visibility $V_0$ at no scattering such that $\xi = 0$ (equivalent to the mean of the modulation function). In a typical imaging setting this would be the visibility of the reference image without the object. In our case, to determine $V_0$ we used the set of mean visibility $V_i$ for each mask acquired at different correlation lengths $\xi_i$ with $i = 1, 2, ..., 30$. We performed a fit for the mean visibility values normalized by visibility at the smallest probed scattering length $\xi_{min}$ for each mask according to 
\begin{equation}
\frac{V_i}{V_{\xi_{min}}}=\exp \left[\sigma t(G(\xi_i)-G(\xi_{min}))\right]
\label{eq:eight},
\end{equation}
where $G(\xi)$ is the projection of autocorrelation function at the correlation length $\xi$ (Eq.~\ref{eq:G_b}). The obtained fitting parameters were plugged into the following equation
\begin{equation}
V_0 = \frac{V_{\xi_i}}{\exp \left[\sigma t(G(\xi_i)-1)\right]}
\end{equation}
to obtain the mean visibility values at no scattering $V_0$ for each mask, which are depicted in Fig~\ref{fig:fig2} (a,b).

We applied a fit for all the data obtained with different mask periods for each mask type using the values $V_0$ for each dataset and achieve fine sampling of the scattering intensity over the range from 90 to 980 nm. The fit was performed using Eq.~\ref{eq:S} and Eq.~\ref{eq:G_b}, and the fitting parameters $\sigma t$, $a$, $H$ were obtained. To check if the fit correctly predicts the value of scattering intensity at correlation lengths larger than 1 $\mu m$, we used the projection image for iHM-50 at the distance of 1120 mm from the detector, corresponding to the correlation length $\xi$ = 1.6 $\mu m$. Note that the extra point acquired for the iHM-50 represents ~1 \% of the data, and its influence on the fitting function can be neglected. As one can see from Fig.~\ref{fig:fig2}(c), the value of the extra data point is well predicted by the fitting function. 

The average pore size $d$ was calculated as a function of the parameters $a$ and $H$  according to Eq.~\ref{eq:d}. The relative pore fraction $\phi_p$ under the spherical pore assumption can be calculated using the total scattering cross-section $\sigma t$ according to the equation \cite{lynch2011interpretation, andersson2008stress}:
\begin{equation}
\sigma t=\frac{3\pi^2}{\lambda^2} d\left|\Delta\chi\right|^2\phi_p \phi_s t
\label{eq:sigma}
\end{equation}
where $|\Delta\chi|$ is the difference in complex refractive index between graphite and air, $d$ is the average pore size,  $\phi_s$ is a relative fraction of solid graphite, and $t$ the sample thickness. For both mask types the value of pore volume fraction is $\phi_p$ = 22 \%. The parameters obtained from the fit and the calculated values of the average pore size and the relative pore fractions are presented in Table~\ref{tab:table1}. The values of average pore size and Hurst exponent can help to estimate the pore size distribution (see Supplemental Material). 
\begin{table*}
\caption{\label{tab:table1}
Parameters derived from the visibility measurements. Macroscopic cross-section $\sigma t$, characteristic parameter $a$ and Hurst exponent $H$ are determined from the fit of experimental data. Average pore size $d$ and relative pore fraction $\phi_p$ were calculated using the values of $a$, $H$ and $\sigma t$.}
\begin{ruledtabular}
\begin{tabular}{cccccc}
\textrm{Mask type}&
\textrm{$\sigma t$}&
\textrm{$a$ (nm)}&
\textrm{$H$}&
\textrm{$d$ (nm) (Eq.~\ref{eq:d})}&
\textrm{$\phi_p$ (Eq.~\ref{eq:sigma})} \\
\colrule
iHM & 0.93 ± 0.02 & 326 ± 15 & 0.58 ± 0.05 & 730 ± 50 & 22 ± 1 \%\\
HM & 0.92 ± 0.03 & 437 ± 32 & 0.58 ± 0.06 & 970 ± 100 & 22 ± 2 \%\\
\end{tabular}
\end{ruledtabular}
\end{table*}

One can see from Table ~\ref{tab:table1}, that the error in fit parameters $a$ and $H$, which are used to calculate the  average pore size $d$ (Eq.~\ref{eq:d}),  for inverted Hartmann masks is noticeably lower than that for the conventional Hartmann masks. The error for average pore size $\Delta d$ was calculated as the error of indirect measurements using partial derivatives of Eq.~\ref{eq:d} (Table~\ref{tab:table1}). The higher $\Delta d$ might be caused by the fact that the area of the inverted Hartmann mask covered with gold is 25 \% of the total field of view of the sample; hence the scattering signal is formed from a larger object area compared to the conventional Hartmann mask. The total amount of scattering centers contributing to the signal is larger, making the obtained results more representative of the bulk structure. The advantage of having a higher signal-to-noise ratio when using the inverted Hartmann mask design for differential phase contrast imaging has been reported before \cite{zakharova2019comparison}.

The visibility map analysis, while being easy and fast to implement, does not provide directional information about the scattering function \cite{pfeiffer2006phase,yashiro2010origin}. One of the advantages of using the Hartmann mask is that it offers periodic modulation in two directions, which enables separation of the horizontal and vertical components of the scattering signal. To profit from that, we applied discrete Fourier transformation to analyze the spatial beam modulation provided by the mask \cite{wen2008spatial,wen2010single}. The spatial frequency spectrum of the Hartmann mask projection contains a strong primary peak around zero spatial frequency and a number of the sharp peaks separated by the $2\pi /P$ distance, where $P$ is the period of the mask. In such a setting, $S_{01}$ and $S_{10}$ are attributed to the first-order Fourier amplitudes in the horizontal and vertical directions, respectively. Since we did not obtain a scattering-free reference image in our measurements (the mask was manufactured directly on the graphite), we analyze only the change in scattering intensity relative to the signal at the smallest correlation length. 

\begin{figure}
\includegraphics[scale=1]{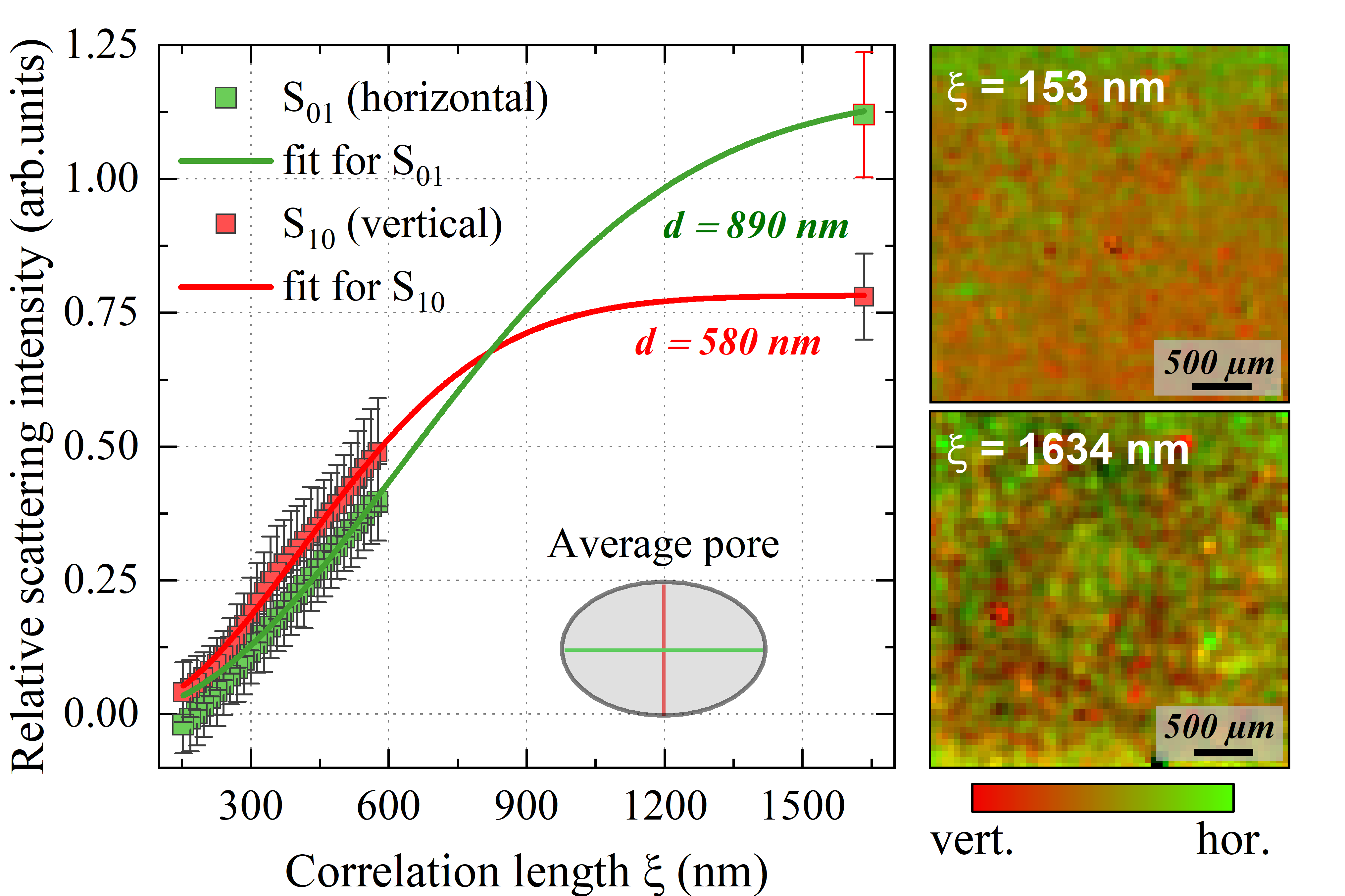}
\caption{\label{fig:fig3} Directional analysis of scattering contrast represented by an overlay of the Fourier amplitudes $S_{01}$ in horizontal (green) and $S_{10}$ in vertical direction (red). The two-dimensional scattering intensity distribution maps for correlation lengths $\xi$ = 153 nm and $\xi$ = 1634 nm are shown on the right as pseudo color images. The inset in the graph shows the shape of an average pore.}
\end{figure}

Since $S_{01}$ and $S_{10}$ are defined for each effective pixel of the imaging system, we obtained the scattering distribution maps in two dimensions for each correlation length $\xi$. Examples of such maps for correlation lengths $\xi$ = 153 nm and $\xi$ = 1634 nm are shown in Fig.~\ref{fig:fig3}. One can see the directional distribution of scatterers in horizontal (green) and vertical (red) directions through the non-even distribution of red and green in the pseudo color images. The mean values of $S_{01}$ and $S_{10}$ for different correlation lengths represented by the data points show that the scattering is mostly isotropic for pores smaller than 580 nm. As the length scale increases up to 1600 nm, the horizontal scattering starts to dominate.

For the data in Fig.~\ref{fig:fig3} we applied the simplified fit according to Eq.~\ref{eq:G_a} with $1 < \alpha < 2$. We determined the characteristic pore size in horizontal and vertical directions to be $d_{hor} = 890 \pm 60$ nm and $d_{vert} = 580 \pm 20$ nm, and the average of the two being 735 nm, which is in agreement with the average pore size obtained by the visibility map analysis. The average pore size in the horizontal direction is larger than in the vertical, indicating the elliptical shape of the characteristic pores (Fig.~\ref{fig:fig3}). Note that relative scattering signal measurements cannot correctly predict the pore fraction and Hurst exponent. 

An important parameter for porous material is its fractal dimension, which indicates how the pores are structured under fractal theory approximation \cite{prostredny2019modelling, adler1993fractal}. The fractal dimension is defined by its Euclidean dimension $D_E$ as well as the value of Hurst exponent. Knowing that the phase boundary parameter $\alpha$ can be determined as $\alpha = H + D_E/2$, we can define the Euclidean dimension of the pore structure in graphite by performing a simplified fit according to the Eq.~\ref{eq:G_a} on the same dataset. The fitting result indicated $\alpha = 1.46 \pm 0.02$ for iHM and $\alpha = 1.52 \pm 0.03$ for HM. From this, we can estimate the Euclidean dimension of the scatterers to be $D_E = 2$. The fractal dimension then is $D = D_E + 1 - H \approx 2.4$, attributed to fractal structures like Apollonian sphere packing ($D = 2.4739465$ \cite{borkovec1994fractal}). 

This paper studied the morphology of bulk pore structure in fine graphite with the scattering contrast available through multi-modal X-ray imaging with Hartmann and inverted Hartmann masks. We scanned the correlation length to study the real-space autocorrelation function of electron density by analyzing the mask visibility reduction. Moreover, we observed the pore size anisotropy by evaluating the relative change in the first-order spatial harmonics using Fourier analysis. 

Based on the presented results, we have determined the pore volume fraction $\phi_p = 22 \%$ and the characteristic pore size $d = 730\pm 50$ nm for measurements with inverted Hartmann mask and $d = 970\pm 100$ nm for conventional Hartmann masks. Both pore fraction and the average pore size values are in close agreement with the values reported for "angstrofine" grade graphite. Considering that the Hurst exponent $H = 0.5$ is characteristic for a perfectly random inhomogeneous solid, the obtained $H = 0.58$ suggests that the distributions of ihnomogeneities in graphite is predominantly random with a slight inclination to being persistent.  The fractal dimension of $D = 2.48$ implies that the pore structure of graphite can be represented by the spheres of different size cotangent to each other \cite{andrade2005apollonian}. Considering the obtained results and calculated errors, we note that the inverted Hartmann mask design may be beneficial for X-ray scattering measurements due to the larger amount of scatterers contributing to the contrast formation.

We expect such a versatile and straightforward technique to impact research devoted to studying complex structures like porous materials, colloids \cite{dallari2021microsecond}, or powders. Apart from the immediate profit for development and characterization of porous catalytic materials, numerous medical applications related to early-stage cancer diagnostics \cite{hunter2006tissue} and lung diseases \cite{taphorn2020grating, helmberger2014quantification} can profit from information on morphology and fractal dimensions of complex interconnected structures.

\begin{acknowledgments}
This work was carried out with the support of KIT light source KARA and Karlsruhe Nano Micro Facility (KNMFi). The authors thank Marcus Zuber and Sabine Bremer for their help during the measurements. The authors acknowledge the funding of the Karlsruhe School of Optics and Photonics (KSOP), associated institution at KIT, and the Karlsruhe House of Young Scientists (KHYS).
\end{acknowledgments}

\bibliography{main}
\end{document}


\title{Bulk morphology of porous materials at sub-micrometer scale studied by multi-modal X-ray imaging \\ - Supplemental Material - }

\author{Margarita Zakharova}
 \email{margarita.zakharova@partner.kit.edu}
\author{Andrey Mikhaylov}
\affiliation{%
 Institute of Microstructure Technology (IMT), Karlsruhe Institute of Technology (KIT), 76344 Eggenstein-Leopoldshafen, Germany
}%

\author{Stefan Reich}%
\affiliation{%
 Institute of Photon Science and Synchrotron Radiation (IPS), Karlsruhe Institute of Technology (KIT), 76344 Eggenstein-Leopoldshafen, Germany
}%
\affiliation{
 Current affiliation: Fraunhofer Institute for High-Speed Dynamics, Ernst-Mach-Institute, EMI, 79104 Freiburg, Germany
}%

\author{Anton Plech}
\affiliation{%
 Institute of Photon Science and Synchrotron Radiation (IPS), Karlsruhe Institute of Technology (KIT), 76344 Eggenstein-Leopoldshafen, Germany
}%

\author{Danays Kunka}
 \email{danays.kunka@kit.edu}
\affiliation{%
 Institute of Microstructure Technology (IMT), Karlsruhe Institute of Technology (KIT), 76344 Eggenstein-Leopoldshafen, Germany
}%

\date{\today}
\maketitle


\onecolumngrid
\section{Graphite characterization}

The wafers purchased from Ohio Carbon Blank, Inc were synthetic graphite labeled "angstrofine" with a porous structure and an average grain size of 1 $\mu m$ as reported by the supplier (supplier identifier EDM-AF5).
The apparent density of the EDM-AF grade graphite material is reported to be 1.8 g/cm\textsuperscript{3}. 
Since the apparent density measurements include the pore volume in the calculation and the theoretical density of graphite is 2.26 g/cm\textsuperscript{3}, the pore fraction of graphite is at least 20 \%. The average pore size for the graphite with an apparent density of 1.8 g/cm\textsuperscript{3} is 750 nm ± 150 nm as observed by mercury porosimetry \cite{Entegris2013}. The distribution of pore sizes for EDM-AF grade graphite is narrow compared to conventional graphite with reported nominal pore size rating from 0.2 to 0.8 $\mu m$ \cite{Entegris2013}. 

\begin{figure}
\includegraphics[scale=0.4]{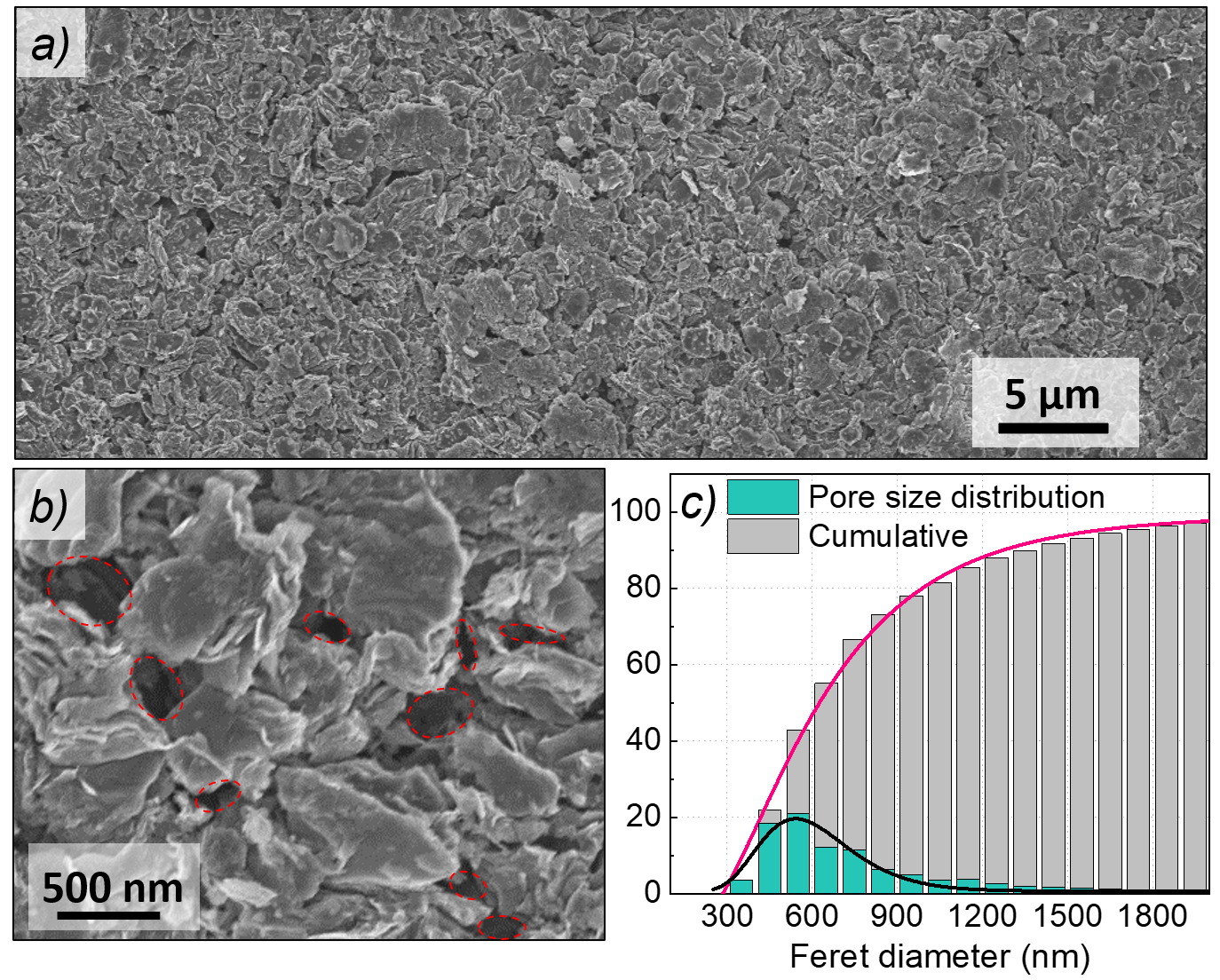}
\caption{\label{fig:figS1}The surface structure of graphite: SEM image of graphite surface (a) with a close-up view (b) outlining the pores in red; c) surface pore size distributions histogram (green) and cumulative pore number (gray) versus Feret diameter based on the SEM image analysis.}
\end{figure}

Based on a set of SEM images (Fig.~\ref{fig:figS1}(a,b)), we performed surface pore size analysis. We identified the pores on the images by thresholding. Then, we estimated their Feret diameter (the longest distance between any two points along the selection boundary) to obtain the pore size distribution histogram (Fig.~\ref{fig:figS1}(c)). The histogram follows a log-normal distribution with the peak at 550 nm. One can see that more than 60 \% of the pores are below 800 nm, which is in agreement with the nominal pore size rating reported for graphite wafers of EDM-AF grade \cite{Entegris2013}. 

\section{Hartmann mask manufacturing}
\begin{figure}
\includegraphics[scale=0.6]{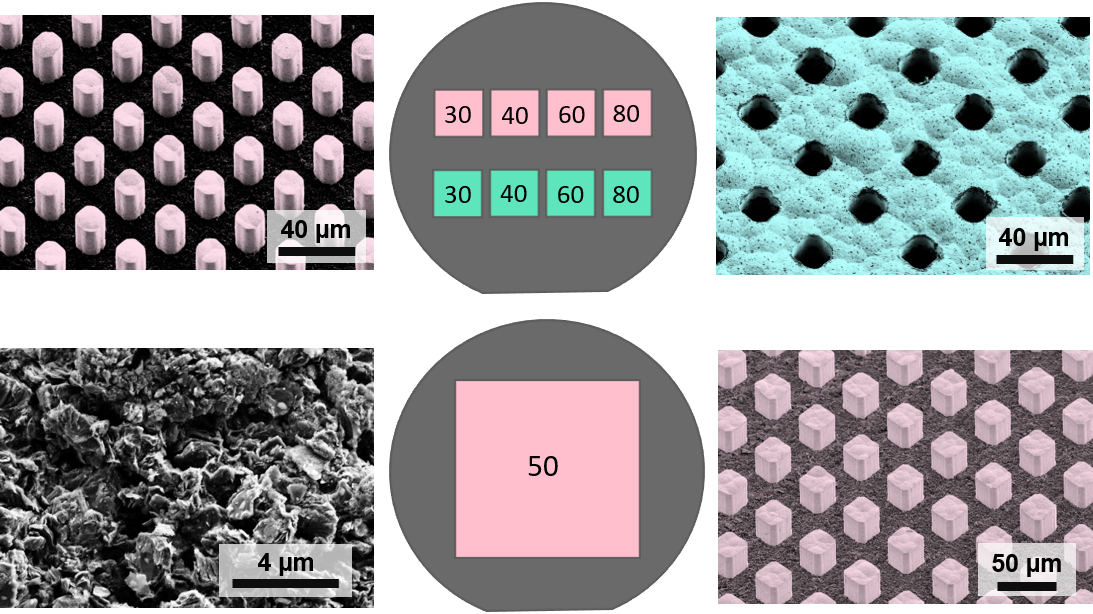}
\caption{\label{fig:figS2}Outline of the wafers used for experiment with SEM images of the masks and the graphite wafer surface. In SEM images the masks were rotated to 45° for better visualization. Inverted Hartmann mask areas are denoted as pink squares and Hartmann masks as green squares with periods 30, 40, 60, 80 $\mu m$. The rows of pillars/holes are alighted parallel to the flat edge of the wafer. On the first wafer, all the areas of the masks were 1 cm\textsuperscript{2}. The inverted Hartmann mask with 50 $\mu m$ period and area of 25 cm\textsuperscript{2} was manufactured on a separate wafer.}
\end{figure}

The separate wafers with a 4-inch diameter were cut out of the large graphite plate. Subsequently, the wafers were polished on both sides and rinsed in isopropanol. The final thickness of the wafers after processing was 500 $\mu m$. 
We followed the technological process of fabricating Hartmann and inverted Hartmann masks that we have reported before \cite{zakharova2019comparison}. The Hartmann masks were manufactured on the graphite wafer in a cleanroom environment using UV lithography, gold electroplating, and resist stripping. Inverted and conventional Hartmann masks of various periods (30, 40, 60, and 80 $\mu m$) with the gold height of about 30 $\mu m$ have been manufactured on the same substrate as indicated in Fig.~\ref{fig:figS2}. Additionally, a large area (5 cm x 5 cm) inverted Hartmann mask was manufactured on a different wafer cut from the same graphite plate. The masks will be further referred to as "Mask type-Period", e.g. iHM-30 for the inverted Hartmann mask of period 30 $\mu m$ and HM-30 for conventional Hartmann mask of period 30 $\mu m$.

\section{Experimental setup}

In the reported method we attribute the decrease in visibility to the ultra small angle X-ray scattering in graphite. However, the imperfections of the setup can also cause a possible decrease in the visibility during sample and optical element movement, because the source size and the source-sample distance are both finite. Large source size (opposite to the point-like source) will cause penumbral blur, and the insufficient distance between the source and the sample will contradict the parallel beam geometry assumption because of the beam divergence. 
We analyzed the setup characteristics to ensure that the assumptions of point-like source and parallel beam are valid for the performed measurements.
\begin{figure}
\includegraphics[scale=0.6]{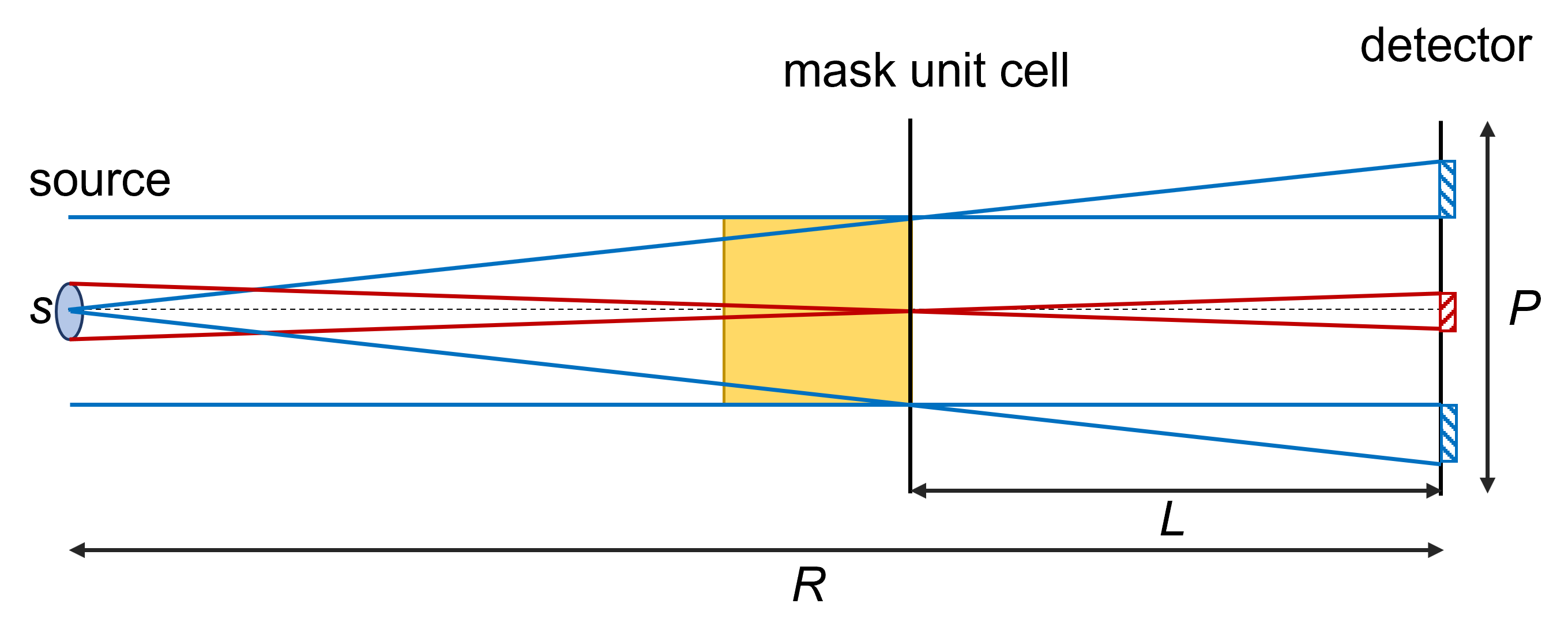}
\caption{\label{fig:figS3}Schematic representation of blur of the mask image induced by the finite source size (red lines) and deviation from the parallel-beam geometry (blue lines) for the horizontal direction (not to scale). The source with the size $s$ is located at the distance $R \approx 30$ m from the detector, the maximum mask-detector distance is $L = 0.395$ m, and the period of the mask is denoted as $P$. Here, the pillar of the inverted Hartmann mask is shown, however, the same considerations are valid for the conventional Hartmann mask.}
\end{figure}
\subsection{Source blurring}
At the IPS imaging cluster of the KIT synchrotron facility (Topo-Tomo beamline), the horizontal source size (800 µm) is larger than the vertical (200 µm). Therefore the blurring caused by the finite source size is stronger in the horizontal direction. The most substantial blur will occur at the largest mask-detector distance; thus, we consider conditions for horizontal direction and the largest mask-detector distance as the limiting factor. The mask-detector distance of 0.395 m was the largest for all masks but iHM-50, for which the maximum mask-detector distance was 1.12 m. The maximum blur of the projection at the detector plane due to the finite source size can be defined as 
\begin{equation}
B_{max} = s_{hor} \cdot \frac{L}{R-L} \approx s_{hor} \cdot \frac{L}{R}
\label{eq:S1},
\end{equation}
where $s_{hor}$ is source size in the horizontal direction, $L$ is object-detector distance, $R$ is source-detector distance (Fig.~\ref{fig:figS3}). 

For the maximum blur to affect the visibility map, it has to be larger than the mask period P (which is the pixel size at the final visibility map image) so that the minimum intensity value is increased, making the visibility value lower. For the masks with periods 30, 40, 60, and 80 µm, the maximum blur is 10 µm, and for the iHM-50 placed at 1.12 m from the detector, the blur is 30 µm (Table~\ref{tab:tableS1}). In both cases, the blur is smaller than the corresponding mask period; thus, it does not significantly affect the mean visibility value. 

\subsection{Beam divergence}
Another important factor to possibly affect the visibility is the beam divergence: the deviation of the rays from the parallel geometry.  The maximum deviation from the parallel-beam projection is observed for the outermost pillar/hole position (Fig.~\ref{fig:figS3}). The deviation can be neglected if it is smaller than the pixel size ($PS$) of the detector. From the scheme in Fig.~\ref{fig:figS3} one can see that the size of the deviation can be defined as 

\begin{equation}
    deviation = \frac{P \cdot L}{R} < PS.
\end{equation}
which leads to the restriction on the mask period
\begin{equation}
    P<\frac{PS \cdot R}{L}
    \label{eq:S2}.
\end{equation}
One can see that for the pixel size $PS = 2.36\ \mu m$, distances $L=0.395\ m$ and $R=30\ m$, the requirement on the mask period is
\begin{equation}
    P < \frac{ 2.36 \cdot 10^{-6} \cdot 30}{0.395} \approx 180\ \mu m.
\end{equation}
This condition is fulfilled for all masks used in the experiment. For the iHM-50, which was placed at the distance $d = 1.12$ m, the same condition holds true, but by a smaller margin:
\begin{equation}
    P<\frac{ 2.36 \cdot 10^{-6} \cdot 30}{1.12} \approx 63\ \mu m.
\end{equation}
Therefore, the divergence of the beam did not significantly affect the mean visibility value acquired in the experiment.

Alternatively, this criteria can be formulated as dimensionless ratio imposing a restriction on the number of pixels per mask period $N$ (sampling of mask projection)
\begin{equation}
    N < R/L
    \label{eq:S3}
\end{equation}
As one can see from the values presented in Table~\ref{tab:tableS1}, the inequality given by Eq.~\ref{eq:S3} holds true for all measurements. 

\subsection{Energy bandwidth and the correlation length error}
The measurements were performed for the energy 17 keV with bandwidth $\Delta E/E \approx 2 \%$, $E = 17 \pm 0.34\ keV$. The correlation length value $\xi$ depends on the energy as:
\begin{equation}
    \xi=\frac{hc}{E} \cdot \frac{L}{P}
\end{equation}
The error of the indirect measurement for $\xi$ will be defined as:
\begin{equation}
    \Delta_\xi = \sqrt{\left(\frac{\partial \xi}{\partial E} \Delta E\right)^2} = \sqrt{\left(-\frac{hc \cdot L}{P\cdot E^2} \Delta E\right)^2}.
\end{equation}
The maximum error due to the bandwidth $\Delta E/E$  is $\Delta_\xi = 33\ nm$  and is reached for the longest mask-detector distance $L=1.12\ m$ and the $P= 50 \ \mu m$. For the distance $L=0.395 \ m$  the highest error $\Delta_Z \approx 20\ nm$ corresponds to the smallest period $P=30\ \mu m$ (Table~\ref{tab:tableS1}). Such small errors do not affect the fitting parameters derived from the data.

\begin{table}
\caption{\label{tab:tableS1}
Setup criteria for plane wave assumption.}
\begin{ruledtabular}
\begin{tabular}{ccccccc}
\textrm{Period ($\mu m$)}&
\textrm{$L$-range (mm)}&
\textrm{$B_{max}$ in experiment ($\mu m$)}&
\textrm{$N$} &
\textrm{$R/L$} &
\textrm{$\xi$-range (nm)} &
\textrm{$\Delta_\xi^{max}$ (nm)}\\
\colrule
30& 95-395 & 10 & 13 & 75 & 231 - 960& 19 \\
40& 95-395 & 10 & 15 & 75& 173 - 720  & 14 \\
50 \footnote{Only iHM}& 95-395, 1120 & 30 & 20 & 26 & 139 - 576, 1634 & 33 \\
60 & 95-395& 10 & 25 & 75 & 115 - 480 & 10 \\
80 & 95-395& 10 & 31 & 75& 87 - 360 & 7 \\
\end{tabular}
\end{ruledtabular}
\end{table}

\section{Simulated pore size distribution}
The obtained values of average pore size and Hurst exponent can help to predict the pore size distribution. Practically important information about the pore size distribution is the peak (mode) of the distribution that indicates the most represented pore size, average (mean) pore size, and the width of the distribution - the range which contains most of the pores.

If we assume that pore sizes $X$ are following a log-normal distribution, the expected value of the $ln(X)$ will be $\mu = ln(d)$, with $d$ being the median of the pore size distribution. This value we obtain as characteristic pore size $d$. In case of log-normal distribution, the median equals a multiplicative mean, which is in agreement with the typical pore structuring: smaller pores cluster and form larger pores, such that the cluster size grows proportionally to the size of the individual pore.

\begin{figure}
\includegraphics[scale=1]{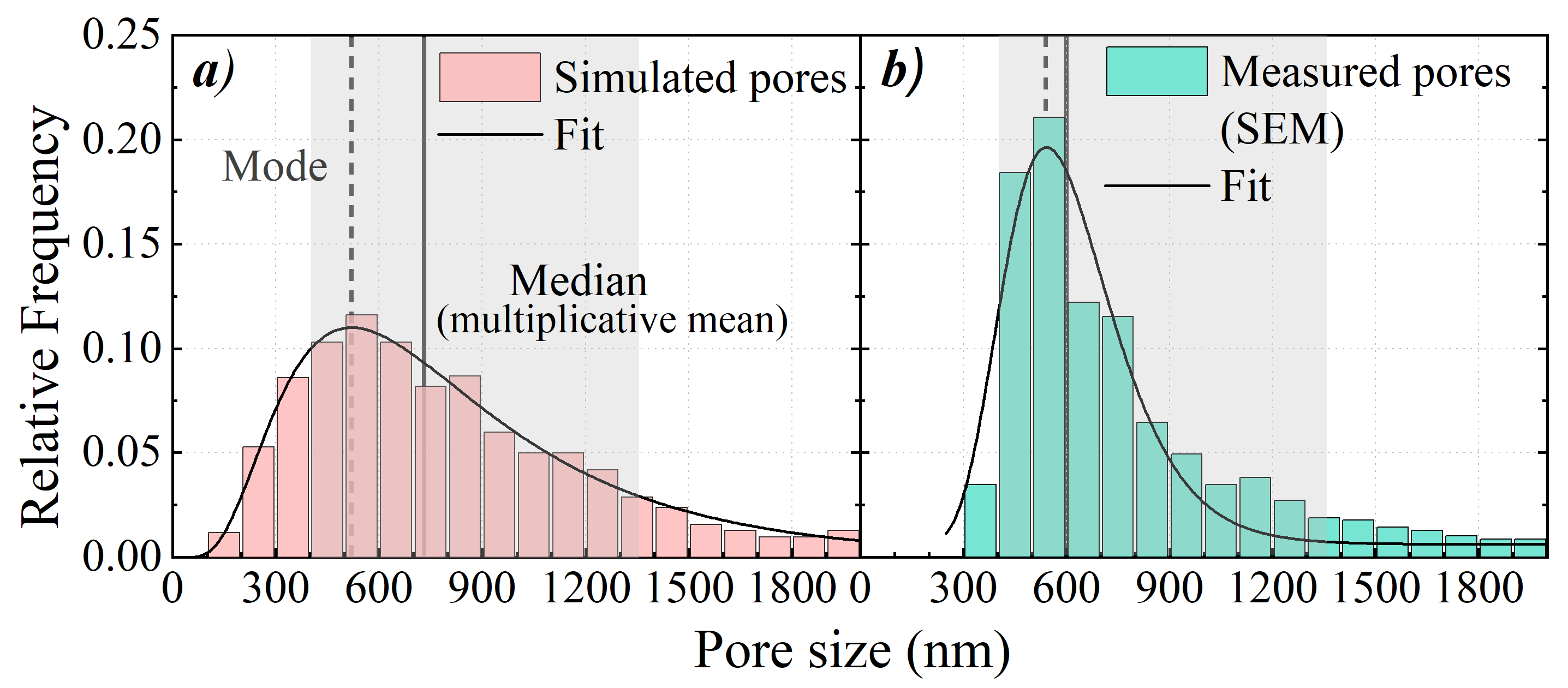}
\caption{\label{fig:figS4}Comparison of the pore size distributions: a) simulated bulk pore size distribution based on the scattering data, b) surface pore size distribution obtained from a set of SEM images. Modes of the distributions are indicated by the dashed lines and the medians by the solid lines. Gray filled area represents the scatter intervals of the distributions containing 2/3 of all pores.}
\end{figure}

The Hurst exponent characterizes the deviation of the electron density distribution from the mean value (or "roughness of the distribution"). If we assume it to serve as an estimate of the standard deviation of the random variable $ln(X)$, the geometric standard deviation factor will be $e^H$. Using the values of avergae pore size $d$ and Hurst exponent $H$, we can evaluate the pore size range containing 2/3 of all pores as the scatter intervals of the distribution from  $d/ e^H = 400$ nm to $d\cdot e^H = 1340$ nm. We can simulate the log-normal distribution of a random variable $X$ based on the data obtained by the scattering contrast (Fig.~\ref{fig:figS4}a). Therefore, the peak of the distribution of pore sizes $X$ is
\begin{equation}
Mode\left|X\right|=\exp{\left[\ln{\left(d\right)-H^2}\right]} = d\cdot e^{-H^2}\approx520\ nm.
\end{equation}

In Fig.~\ref{fig:figS4}, we compare the simulated bulk pore size distribution to the surface pore size distribution obtained from the set of SEM images. Note that the distribution obtained from SEM images is not comprehensive and is only valid for a restricted field of view (hence with limited pore size statistics) on the surface of graphite. Nevertheless, it is a valuable benchmark to see if the simulated distribution is realistic. Fig.~\ref{fig:figS4} shows that, although the distributions differ in shape, the mode of the simulated distribution (dashed line at 520 nm in Fig.~\ref{fig:figS4}(a)) is close to the mode of the surface pore size distribution (shown as the dashed line  at 540 nm  in Fig.~\ref{fig:figS4}(b)). The median of the simulated distribution (730 nm) is different from the median of the measured one (600 nm); we can see that the fit for the measured SEM distribution does not approximate the larger pores in the range from 900 nm to 1500 nm. Accounting for these pore clusters would make the distribution wider, shifting the median towards larger values.

\bibliography{supp}